\documentclass[%
 reprint,
showpacs,preprintnumbers,
 amsmath,amssymb,
 aps,
]{revtex4-1}
\usepackage{graphicx}
\usepackage{dcolumn}
\usepackage{bm}
\usepackage{epstopdf}

\begin{document}

\preprint{APS/123-QED}

\title{Centrality dependence of charged-hadron pseudorapidity distributions in PbPb collisions at LHC energies in the RDM}

\author{David M. R\"ohrscheid}
\author{Georg Wolschin}%
 \email{wolschin@uni-hd.de}
\affiliation{%
 Institut f{\"ur} Theoretische 
Physik
der Universit{\"a}t Heidelberg, Philosophenweg 16, D-69120 Heidelberg, Germany, EU\\
}%


\date{\today}

\begin{abstract}
The centrality dependence of charged-hadron production in relativistic heavy-ion 
collisions at LHC energies is investigated in a nonequilibrium-statistical relativistic diffusion model (RDM) 
with three sources. Theoretical pseudorapidity distributions are compared with 
preliminary PbPb ALICE data at LHC energy of $\sqrt{s_{NN}}$ = 2.76 TeV for seven centralities, and a previous RDM prediction for this energy is confronted with the data. Refined predictions for 5.52 TeV PbPb based on RDM parameter extrapolations are presented.

\end{abstract}

\maketitle
\section{\label{sec:intro}Introduction\protect} 

First results of charged-hadron production in PbPb collisions   
 at LHC energies of $\sqrt {s_{NN}}$ = 2.76 TeV \cite{aa10,aamo10}
have recently been analyzed in a nonequilibrium-statistical relativistic diffusion model (RDM) \cite{gw11}.  So far, the RDM calculations have been compared to central collision data.  Now, preliminary ALICE results from the 2010 LHC run at different centralities, and for a large range of pseudorapidities $-3.7<\eta<5.1$ have become available \cite{to11} which we use in this work    for a comparison with our RDM calculations, and for a determination of the corresponding parameters,
which had in \cite{gw11} only been extrapolated to LHC energies. 

In the RDM, the underlying three sources for particle production can be traced back to a midrapidity source resulting from gluon-gluon collisions, and two forward-centered sources arising from valence quark-gluon collisions.
The particle production sources are broadenend in time through nonequilibrium-statistical processes such as collisions and particle creations, which are described based on a Fokker-Planck type transport equation \cite{wol99,wols99,wol07}. The additional broadening of the distribution functions due to collective expansion then leads to an effective diffusion coefficient \cite{wol06}. 

Even in the presence of expansion, drift and diffusion coefficients are related through a dissipation-fluctuation theorem in the nonlinear version \cite{alb00,ryb03,wol03} of the model where a nonextensivity parameter $q$ implicitly accounts for the effect of long-range interactions that are considered to cause the collective behaviour.
In the linear model, expansion is treated explicitly \cite{wol06}. Here the relaxation time (or the drift coefficient) and the diffusion coefficient are considered as independent parameters.  

An incoherent superposition of the
three sources at the interaction time -- where the integration of the transport equation stops --
yields good agreement with charged-particle pseudorapidity distributions at RHIC energies.
It has been shown in \cite{biy04,wob06,kw07} within the RDM that at RHIC energies of 0.13 TeV (0.2 TeV) the midrapidity source generates about  13 \% (26 \%) of the produced particles in a 0--6\% central AuAu collision, whereas the bulk of the particles is still produced in the two fragmentation sources. At SPS, and low RHIC energies of 19.6 GeV the effect of the midrapidity source is negligible \cite{kw07}.

At sufficiently high energies (RHIC at 0.13 and 0.2 TeV), the effect of the Jacobian transformation from rapidity to pseudorapidity space together with the superposition of the sources produces a slight midrapidity dip in the charged-hadron pseudorapidity distribution \cite{kw07}. At LHC energies, the effect of the Jacobian for a given particle mass tends to be smaller due to the higher average transverse momenta \cite{gw11}, but the dip in the preliminary ALICE data is clearly more pronounced \cite{to11} than at RHIC energies. If confirmed, this would be a strong hint towards the significance of the superposition of the underlying sources, with the fragmentation sources at LHC energies moving further apart in pseudorapidity space due to the higher beam rapidity.

The relativistic diffusion model has previously also been applied to $pp$ collisions at RHIC and LHC energies \cite{gwo11}. Although transport phenomena are not expected to be fully developed here due to the small transverse size of the system, the RDM yields reasonable results for pseudorapidity distributions of produced charged hadrons in the $\eta-$range where data are available, and provides predictions for large values of $\eta$. 

In a recent paper \cite{tot12},
the TOTEM collaboration has indeed reported first experimental $pp$ results at 7 TeV and large $|\eta|$. Measured values for $dN/d\eta$ of charged hadrons range from 3.84$\pm$0.01(stat)$\pm$0.37(syst) at $|\eta| = 5.375$ to 2.38$\pm$ 0.01(stat)$\pm$0.21(syst) at $|\eta| = 6.37$. The data account for about 95\% of the total inelastic cross section. A corresponding RDM prediction as shown in Fig. 4 of \cite{gwo11}  normalized to 7 TeV CMS NSD $pp$ data at midrapidity (which are below inelastic results) turns out to be slightly below the TOTEM values at large $|\eta|$. In contrast to MC predictions, it has the correct slope. 

Our model is also suited for asymmetric systems such as $d$+Au and $p$+Pb, which are more sensitive to the details of the distribution functions. At 0.2 TeV we found a midrapidity source containing 19 \% of the produced particles for 0--20\% central $d$+Au collisions \cite{wobi06}. Particle creation from a gluon-dominated midrapidity source, incoherently added to the sources related to the valence part of the nucleons, had also been proposed by Bialas and Czyz \cite{bia05}. Another three-sources model has recently been used to study the influence of the details of hadronization and freeze-out on the constituent quark-number scaling in the elliptic flow \cite{tsch11}.
 
There exist many complementary models which assume a central source such as the dual parton model \cite{cap82,cap94}, or the quark-gluon string model \cite{kai03}. Models with specific applications to heavy-ion collisions at LHC energies have been summarized in \cite{arm08,arme09}. Approaches based on gluon saturation such as \cite{nar05,alb07} usually do not account explicitly for the contributions of the fragmentation sources. The RDM provides a simple and transparent analytical framework to investigate the interplay of central and fragmentation sources in heavy-ion collisions at relativistic energies.

Within the RDM, we investigate in this work the centrality and energy dependence of the three sources for particle production in collisions of symmetric systems at RHIC and LHC energies in direct comparison with data, and provide a prediction for central collisions at maximum LHC energies.

The energy range considered here for the three-sources model covers RHIC energies of $\sqrt {s_{NN}}$ = 0.019, 0.062, 0.13 and 0.2 TeV in 
AuAu collisions, the presently accessible LHC energy of 2.76 TeV in PbPb collisions, and the maximum LHC energy of 5.52 TeV.

The ingredients of the model are briefly reconsidered in the following section, the calculations of pseudorapidity distributions of charged hadrons at RHIC and LHC energies are discussed in the third section, results and the determination of the RDM parameters depending on energy and centrality are performed in the fourth section. Conclusions are drawn in the fifth section.

\section{Relativistic Diffusion Model}


\begin{figure}
\centering
\includegraphics[width=8.8cm]{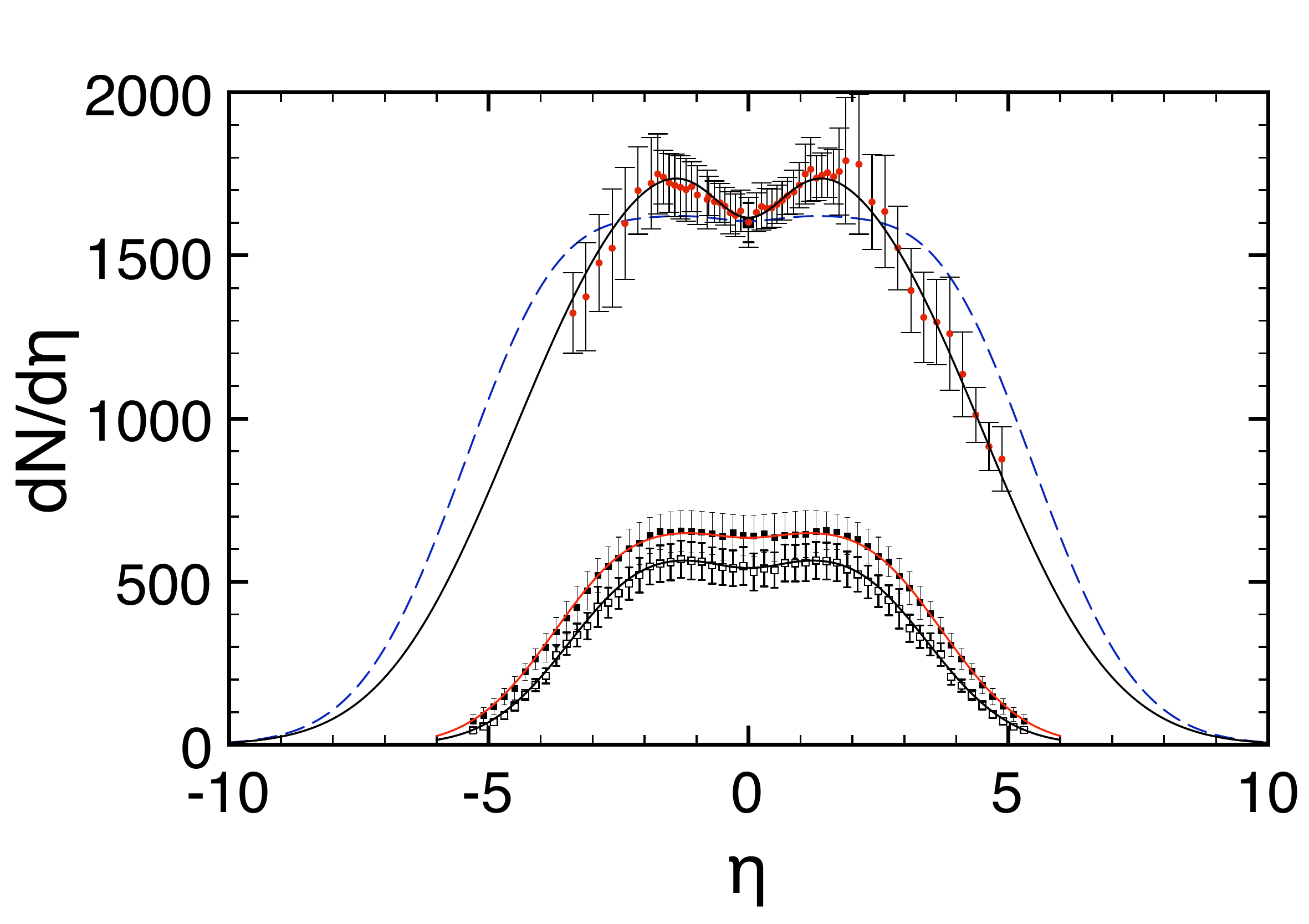}
\caption{
\label{fig1}
(Color online) The predicted (dashed \cite{gw11}) RDM  pseudorapidity distribution function for charged hadrons in 0--5\% central PbPb collisions
at LHC energies of 2.76 TeV is shown in the upper part of the figure, with the mid-rapidity value adjusted to the ALICE data point \cite{aamo10}, and RDM parameters as in \cite{gw11}. The solid curve is a $\chi^2$-minimization based on the three-sources RDM with respect to the preliminary ALICE data from \cite{to11}
that takes the limiting fragmentation hypothesis into account, see text. Error bars have been symmetrized using the larger branches. The corresponding 
RDM parameters are given in Table~\ref{tab1}, and Fig.~2. 
In the lower part of the figure,
calculated pseudorapidity distributions of 
produced charged particles from
AuAu collisions (bottom) at $\sqrt{s_{NN}}$ =  0.13 and 0.2 TeV for
0--6\% central collisions with PHOBOS data \cite{bb01,bb03,alv11}
 are shown for comparison, see \cite{gw11}. 
}
\end{figure}

In the Relativistic Diffusion Model, the time evolution of the distribution
functions is governed by a Fokker-Planck
equation (FPE) in rapidity space
\cite{wol07} (and references therein)
\begin{equation}
\frac{\partial}{\partial t} R(y,t)=-\frac{\partial}
{\partial y}\Bigl[J(y) R(y,t)\Bigr]+
D_{y}\frac{\partial^2}{\partial y^2}[R(y,t)]^{2-q} 
\label{fpenl}
\end{equation}
with the rapidity $y=0.5\cdot \ln((E+p)/(E-p))$. The beam rapidity can also be written as 
$y_{\text{beam}}=\mp y_{max}=\mp \ln(\sqrt{s_{NN}}/m_{p})$.
The rapidity diffusion coefficient $D_{y}$ that contains the
microscopic physics accounts for the broadening of the
rapidity distributions.

The drift $J(y)$ determines the shift of the mean rapidities
towards the central value, and linear and nonlinear 
forms have been discussed \cite{alb00,ryb03,wol07}.
In the diffusion term, a nonlinearity may enter for a nonextensivity coefficient \cite{tsa96}
$1 < q < 1.5$ that implicitly accounts for long-range interactions causing collective expansion.
In normal Boltzmann statistics one has
$q = 1$, and the diffusion term is linear \cite{gw04} as in Brownian motion.

The rapidity distribution of produced particles
is calculated from an incoherent superposition of the beam-like fragmentation 
components at larger rapidities arising mostly from valence quark-gluon interactions, and a
component centered at midrapidity that is essentially
due to gluon-gluon collisions. All three distributions are broadened in rapidity space
as a consequence of diffusion-like processes, and due to collective expansion.

Here we use the standard linear FPE that corresponds to $q = 1$ and a
 linear drift function
 \begin{equation}
J(y)=(y_{eq}- y)/\tau_{y}
\label{dri}
\end{equation}
with the rapidity relaxation time $\tau_{y}$, and the equilibrium 
value $y_{eq}$ of the rapidity. For symmetric systems as investigated in the present work,
$y_{eq} = 0$, whereas $y_{eq}$ differs from 0 in asymmetric systems \cite{wob06}.

The linear formulation of the transport equation corresponds to
the so-called Uhlenbeck-Ornstein \cite{uhl30} process, applied to the
relativistic invariant rapidity for the three components  
$R_{k}(y,t)$ ($k$=1,2,3) of the distribution function
in rapidity space
\begin{equation}
\frac{\partial}{\partial t} R_{k}(y,t) =
-\frac{1}{\tau_{y}}\frac{\partial}
{\partial y}\Bigl[(y_{eq}-y)\cdot R_{k}(y,t)\Bigr]
+D_{y}^{k} \frac{\partial^2}{\partial y^2}
R_{k}(y,t).
\label{fpe}
\end{equation}

In the linear case, a superposition of the distribution
functions \cite{wol99,wol03} using the initial conditions
$R_{1,2}(y,t=0)=\delta(y\pm y_{max})$
with the absolute value of the beam rapidities 
$y_{max}$, and $R_{3}(y,t=0)=\delta(y-y_{eq})$  
yields the exact solution. The mean values 
are derived analytically from the moments 
equations as
\begin{equation}
<y_{1,2}(t)>=y_{eq}[1-\exp(-t/\tau_{y})] \mp y_{max}\exp{(-t/\tau_{y})}
\label{mean}
\end{equation}
for the sources (1) and (2) with the absolute value of the beam rapidity $y_{max}$,
and $y_{eq}$ for the local equilibrium source which is equal to zero only for
symmetric systems. For asymmetric systems, the midrapidity source is moving
\cite{wob06}, and the superposition of the sources is very sensitive to
the values of the model parameters.

Both mean values $<y_{1,2}>$ would attain y$_{eq}$ 
for t$\rightarrow \infty$, whereas for short times they remain between 
beam and equilibrium values. The variances are
\begin{equation}
\sigma_{k}^{2}(t)=D_{y}^{k}\tau_{y}[1-\exp(-2t/\tau_{y})],
\label{var}
\end{equation}
and the corresponding FWHM-values
are obtained from 
$\Gamma_k=\sqrt{8\ln2}\cdot \sigma_k$ since the partial distribution functions are Gaussians in rapidity space
(but not in pseudorapidity space).

For symmetric systems, the midrapidity source $R_3(y,t)\\ 
= R_{gg}(y,t)$ has mean value zero. Since the width approaches equilibrium twice as fast as the mean value, the central source comes close to thermal equilibrium with respect to the variable rapidity during the interaction time
$\tau_{int}$. Full equilibrium as determined by the temperature would be reached for $\tau_{int}/\tau_y \gg 1$. The fragmentation sources do not reach $<y_{1,2}>=0$ during the interaction time and hence, remain far from thermal distributions in rapidity space, and do not fully equilibrate with the central source.

The charged-particle distribution in rapidity space is obtained
as incoherent 
superposition of nonequilibrium and central (``equilibrium") solutions of
 (\ref{fpe}) 
\begin{eqnarray}
    \lefteqn{
\frac{dN_{ch}(y,t=\tau_{int})}{dy}=N_{ch}^{1}R_{1}(y,\tau_{int})}\nonumber\\&&
\qquad\qquad +N_{ch}^{2}R_{2}(y,\tau_{int})
+N_{ch}^{gg}R_{gg}(y,\tau_{int})
\label{normloc1}
\end{eqnarray}
with the interaction time $\tau_{int}$ (total integration time of the
differential equation).

\section{Pseudorapidity distributions}
To calculate pseudorapidity distributions which depend only on the scattering angle $\theta$, we convert the results from rapidity to $\eta$ space,
$\eta=-$ln[tan($\theta / 2)]$,  through 
\begin{equation}
\frac{dN}{d\eta}=\frac{dN}{dy}\frac{dy}{d\eta}=
J(\eta, \langle m \rangle/ p_{T})\frac{dN}{dy} 
\label{deta}
\end{equation}
using the Jacobian
\begin{eqnarray}
\lefteqn{J(\eta, \langle m \rangle/p_{T})=\cosh({\eta})\cdot }
\nonumber\\&&
\qquad\qquad[1+(\langle m \rangle/ p_{T})^{2}
+\sinh^{2}(\eta)]^{-1/2}.
\label{jac}
\end{eqnarray}

\begin{figure}
\begin{center}
\includegraphics[width=8.6cm]{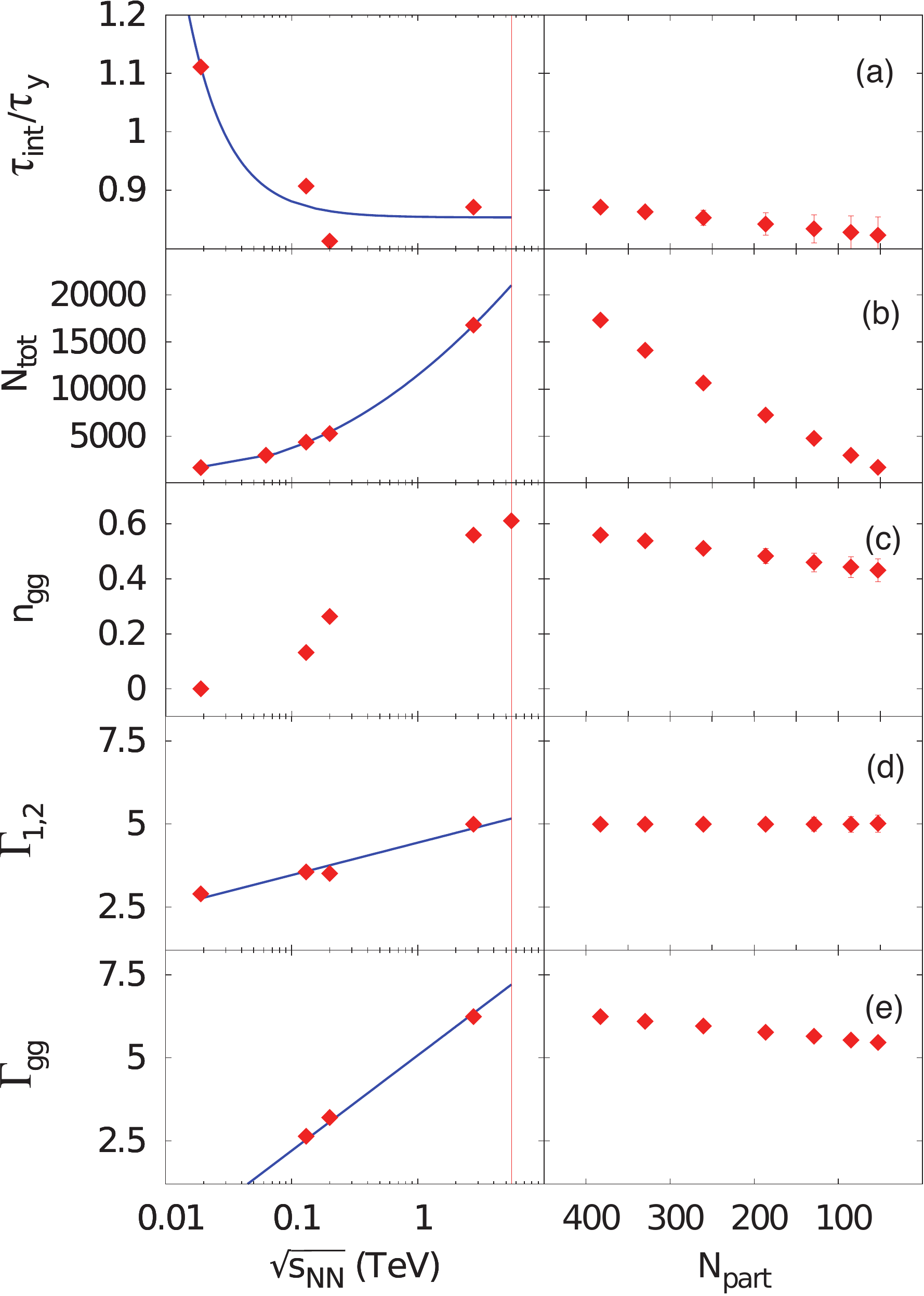}
\caption{\label{fig2}(Color online) Left column: Dependence of the diffusion-model parameters 
for heavy systems
(central AuAu at RHIC energies, central PbPb at LHC energies) on the center-of-mass 
energy $\sqrt{s_{NN}}$ (in GeV) as determined in this work, from top to bottom: (a) Quotient of interaction time and relaxation 
time; (b) total number of produced charged hadrons; (c) fraction of charged hadrons in the central source; 
(d) width (FWHM) of the peripheral sources including collective expansion; (e) width of the midrapidity source.
Results refer to charged-hadron pseudorapidity distributions,
with extrapolations to the highest LHC energy of 5.52 TeV, vertical line. The diamonds are fit values at RHIC energies of 19.6, 62.4, 130 and 200 GeV, and at the LHC energy of 2.76 TeV. Right column:
Corresponding parameter dependencies on the number of participants $N_{\text{part}}$ in PbPb collisions at 
$\sqrt{s_{NN}}$ = 2.76 TeV for seven centralities.}
\end{center}
\end{figure}


The Jacobian depends on both the particle mass $m$ (or the mean particle mass $\langle m \rangle$ if particle identification is not available), and the transverse momentum $p_T$ through the term $(m/p_T)^2$. However, when comparing with measured $dN/d\eta$ distributions of produced charged hadrons, particle identification in rapidity space is not usually available, and one has to resort to an approximate expression for the Jacobian. In previous works (such as \cite{wob06}) we had approximated the mean mass by the pion mass (or a slightly higher value \cite{wbm08}) since most of the produced hadrons are pions, and $p_T$ by the mean transverse momentum $<p_T>$ of pions at midrapidity, which is about 0.45 GeV/$c$ at 0.2 TeV center-of mass energy, and 0.5 GeV/$c$ at 2.76 TeV. 

Here we also consider the effect of more massive produced mesons where the effect of the Jacobian is more pronounced. At LHC energies of 2.76 TeV in PbPb, preliminary $p_T-$spectra for identified $\pi^-, K^-$ and antiprotons are available \cite{pr11} as functions of centrality. Since the particle to antiparticle ratios are very close to one at this energy \cite{pr11}, we take identical spectra for the corresponding antiparticles, and parametrize the $p_T-$dependence with a suitable functional form \cite{dmr12}. The relative particle abundances
in central (0-5\%) PbPb collisions are 83\% for pions, 13\% for kaons, and 4\% for protons, with the pion fraction increasing to 84\% for more peripheral (50-60\%) collisions. The corresponding
$<p_T>-$values calculated from the spectra in central (0-5\%) collisions are 0.51, 0.86 and 1.37 GeV/$c$ for $\pi, K$ and $p$ with the $<p_T>-$value for pions decreasing to 0.47 GeV/$c$ for peripheral (50-60\%) collisions.

To obtain the Jacobian $J = dy/d\eta$ at midrapidity $y=\eta=0$ for the total charged-hadron distribution in a given centrality class, we add the resulting multiplicity densities for the three particle species considered here, and their antiparticles, to obtain $J_{y=0}=(dN/d\eta|_{\eta=0})/(dN/dy|_{y=0})$. For 2.76 TeV PbPb central collisions, this yields $J_{y=0}=0.856$, which is smaller than the result obtained from the analytical expression for the Jacobian (\ref{jac}) calculated with $\langle m \rangle=m_{\pi}$, and the mean transverse momentum. 

This indicates that one should consider the full $p_T-$distributions 
for all charged hadrons when calculating the Jacobian. Since the result is, however, approximately equivalent to the use of  effective values for the mass $<m_\text{eff}>$ and the transverse momentum $<p_{T,\text{eff}}>$ in (\ref{jac}), we choose $<m_{\text{eff}}>=m_{\pi}$, and determine $<p_{T,\text{eff}}>$ such that $J_{y=0}$ for the total distribution of charged hadrons is exactly reproduced. Since the effect of the Jacobian transformation is most pronounced at midrapidity, this minimizes possible deviations from the correct values.
For each centrality class, this yields
\begin{equation}
<p_{T,\text{eff}}>=m_{\pi}J_{y=0}/\sqrt{1-J_{y=0}^2} .
\end{equation}
For central (0-5\%) PbPb collisions at 2.76 TeV, this amounts to $<p_{T,\text{eff}}>=0.21$ GeV/$c$, and for
 50-60\% centrality $<p_{T,\text{eff}}>=0.19$ GeV/$c$. With these effective mean transverse momenta, the corresponding Jacobians for each centrality class are calculated, and the full pseudorapidity distribution functions for produced charged hadrons are obtained in the three-sources model according to (\ref{normloc1}),(\ref{jac}). The RDM parameters are then optimized in a $\chi^2-$minimization with respect to the available data. We show the results in the next section (Figs. 3,4).

\begin{figure}
\begin{center}%
\includegraphics[width=9.6cm]{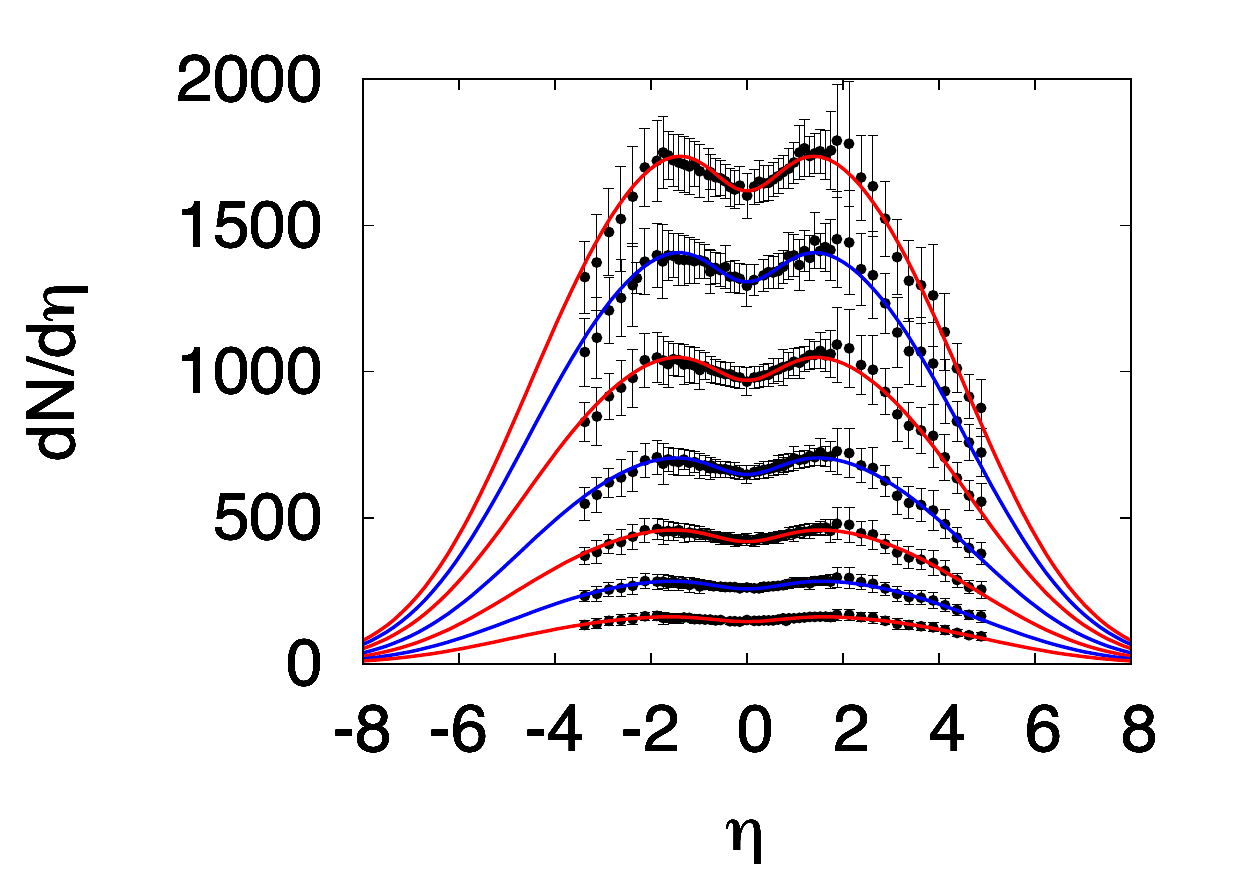}
\caption{\label{fig3} (Color online) Pseudorapidity distributions for produced charged hadrons in 2.76 TeV PbPb collisions as functions of centrality, from bottom to top: 50--60\%, 40--50\%, 30--40\%, 20--30\%, 10--20\%, 5--10\%, 0--5\%. Calculated RDM distributions (solid curves) have been optimized in $\chi^2$-fits with respect to the preliminary ALICE data from \cite{to11}, and using the limiting fragmentation scaling hypothesis in the region of large rapidities where no data are available. The centrality-dependent parameter values are as shown in the rhs column of Fig. 2. }

\end{center}
\end{figure}

\begin{figure}
\begin{center}
\vspace{.25cm}
\includegraphics[width=8.8cm]{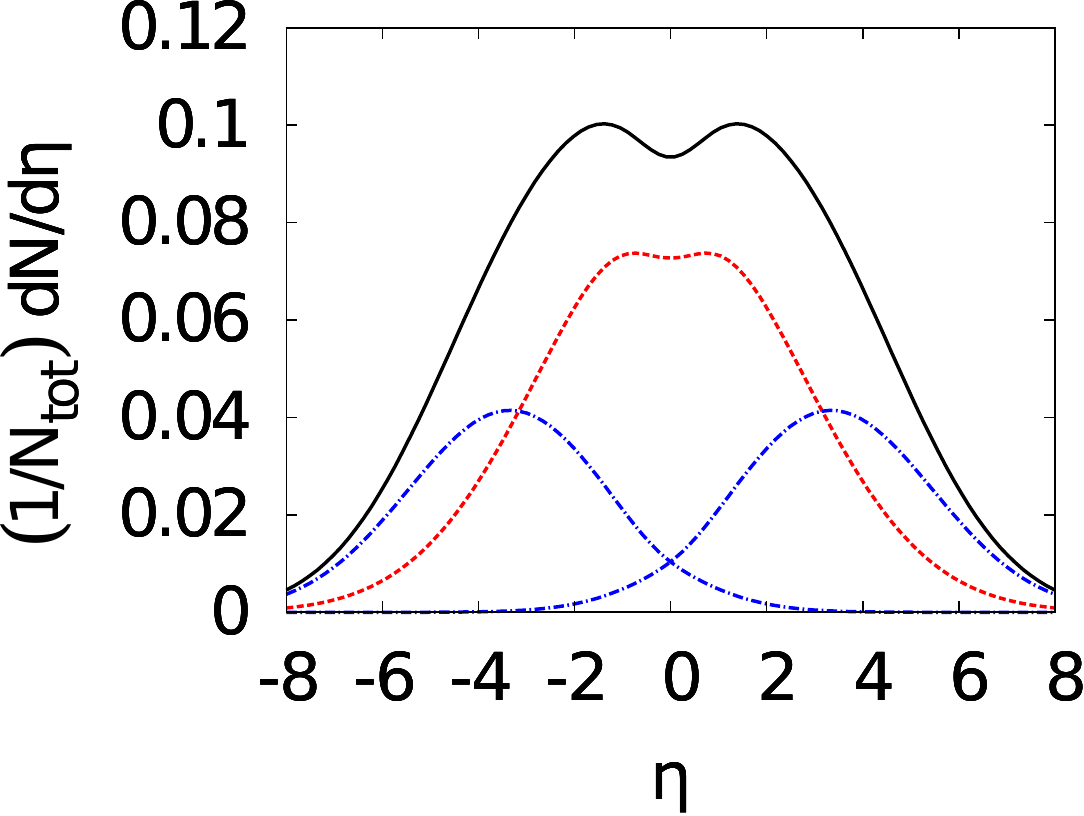}
\caption{\label{fig4}(Color online) Pseudorapidity distribution of charged hadrons in 0--5\% central PbPb collisions at LHC energies of \(\sqrt {s_{NN}}\) =  2.76 TeV, with the underlying theoretical distributions.
The result (solid curve) is based on the interplay of central (gluon--gluon, dashed)  and peripheral (valence quarks -- gluon, dash-dotted) distribution functions. At LHC energies, the midrapidity value is mostly determined by particle production from gluon--gluon collisions.} 
\end{center}
\end{figure}

\begin{figure}
\begin{center}
\includegraphics[width=8.8cm]{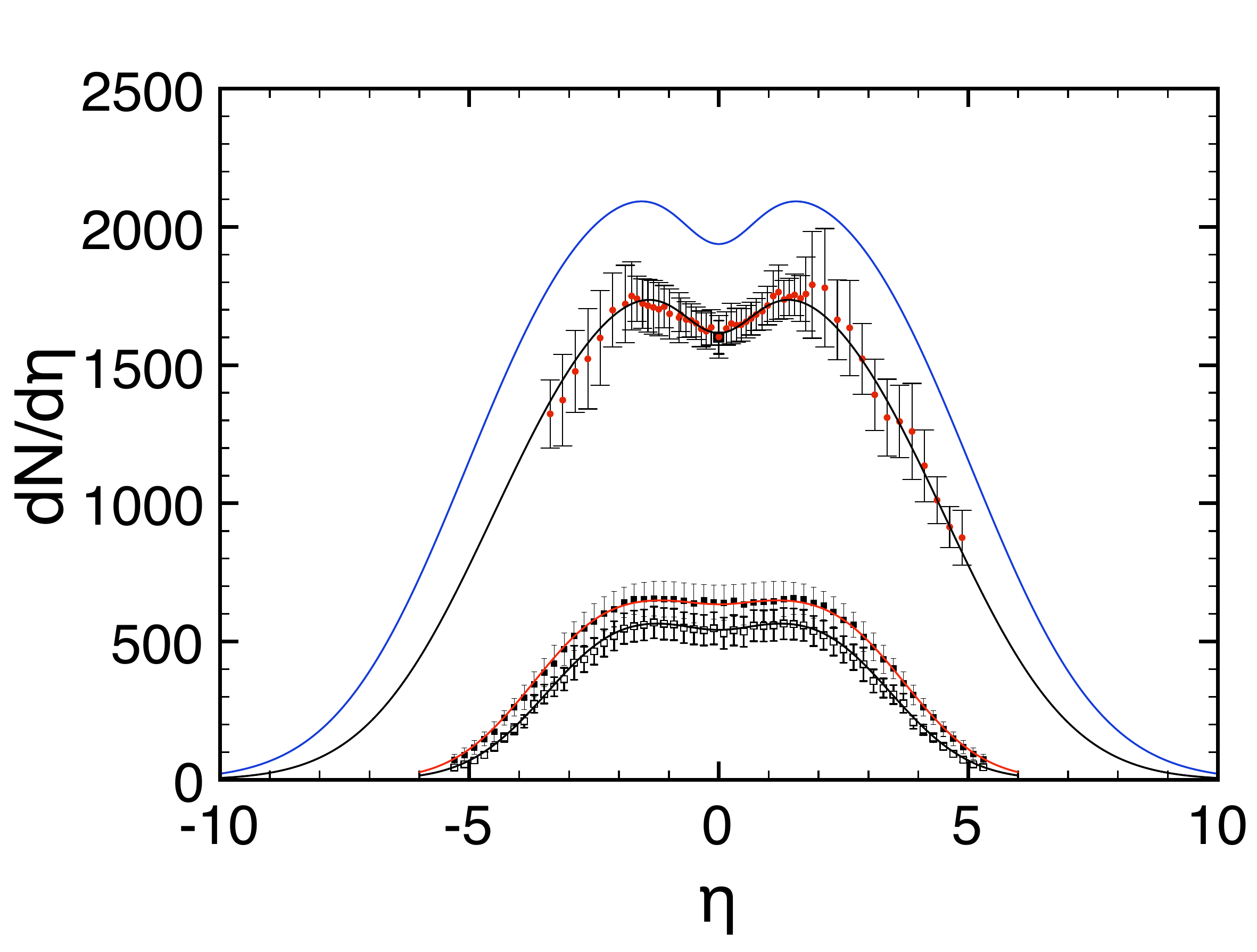}
\caption{\label{fig5}(Color online) Predicted pseudorapidity distribution of charged hadrons in 0--5\% central PbPb collisions at LHC energies of \(\sqrt {s_{NN}}\) = 5.52 TeV with extrapolated RDM-parameters from Table 1 (upper curve). The curve below refers to the $\chi^2-$optimization of the RDM distribution functions with respect to the preliminary ALICE data \cite{to11} in central collisions at 2.76 TeVPbPb, as in Fig. 1.
Results for 0.13 and 0.2 TeV central AuAu are shown for comparison (lower curves, as in Fig. 1).
}
\end{center}
\end{figure}

\begin{table*}
\begin{center}
\caption{\label{tab1}Three-sources RDM-parameters $\tau_{int}/\tau_{y}$, $\Gamma_{1,2},  \Gamma_{gg}$, and $n_{gg}$. $N_{ch}^{tot}$ is the total number of charged particles, $n_{gg}$ the fraction of particles produced in the central source. 
Results for $<y_{1,2}>$ are calculated from $y_{\text{beam}}$ and $\tau_{int}/\tau_{y}$. 
Values are shown
for 0--5\% PbPb at LHC energies of 2.76 and 5.52 TeV in the lower two lines,  with results at 2.76 TeV from a $\chi^2$-minimization with respect to the preliminary ALICE data \cite{to11}, and using limited fragmentation as constraint. See Fig.~\ref{fig2} and text for the evolution of the parameters from RHIC to LHC energies.
Corresponding parameters
for 0--6\% AuAu at RHIC energies are given for comparison in the upper four lines based on recent PHOBOS results \cite{alv11}.
Parameters at 5.52 TeV denoted by * are extrapolated according to  Fig.~\ref{fig2}.  
Experimental midrapidity values (last column) are from PHOBOS \cite{alv11} for $|\eta| < 1$, 0-6\% at RHIC energies and from ALICE \cite{aamo10}  for $|\eta| < 0.5$, 0-5\% at
2.76 TeV.} 
\vspace{.3cm}
\begin{tabular}{lllllllcr}
\hline\\
$\sqrt{s_{NN}} $&$y_{\text{beam}}$& $\tau_{int}/\tau_{y}$&$<y_{1,2}>$&$\Gamma_{1,2}$&$\Gamma_{gg}$&$N_{ch}^{tot}$&$n_{gg}$&$\frac{dN}{d\eta}|_{\eta \simeq 0}$\\
   (TeV)\\
\hline\\
 0.019&$\mp 3.04$&0.97&$\mp 1.16$& 2.83&0&1704&-&314$\pm 23$\cite{alv11}\\
     0.062&$\mp 4.20$&0.89&$\mp 1.72$& 3.24&2.05&3003&0.07&463$\pm 34$\cite{alv11}\\  
  0.13&$\mp 4.93$&0.89&$\mp 2.02$& 3.43&2.46&4398&0.13&579$\pm 23$\cite{alv11}\\
 0.20&$\mp 5.36$&0.82&$\mp 2.40$& 3.48&3.28&5315&0.26&655$\pm 49$ \cite{alv11}\\
 2.76&$\mp 7.99$&0.87& $\mp 3.34$&4.99&6.24&17327&0.56&1601$\pm 60$ \cite{aamo10}\\
  5.52&$\mp 8.68$&0.85*& $\mp 3.70$&5.16*&7.21*&22792*&0.61*&1940*\\\\
\hline 
\end{tabular}
\end{center}
\end{table*}

At RHIC energies, we had investigated the
 dependencies of the diffusion-model parameters on incident energy, mass and centrality 
 in \cite{wob06,kw07,kwo07}. In particular, the centrality dependence seen in the RHIC data is exactly reproduced in the three-sources model \cite{wob06,kw07}. The parameters are shown in Table~\ref{tab1} as functions of the c.m. energy in central collisions of AuAu, and in central PbPb at LHC energies. 

A prediction for central PbPb at 2.76 TeV with extrapolated RDM parameters had been presented in \cite{gw11}, with the midrapidity point adjusted to the ALICE data \cite{aamo10}. It is shown  by the dashed curve in Fig.~\ref{fig1}. Evidently, the predicted fragmentation-peak position is too far from midrapidity compared to the preliminary data \cite{to11}, and the experimental midrapidity dip is more pronounced than in the prediction.

Although these features may be somewhat modified once final LHC data with better statistics become available, we have performed a $\chi^2-$optimization of the three-sources model solutions with respect to the preliminary ALICE data in order to determine the modification of the parameters from the prediction.

\section{Results and RDM-parameters}

A determination of the RDM parameter for PbPb at 2.76 TeV through a straightforward $\chi^2-$minimization with respect to the preliminary ALICE data does not yield satisfactory results
due to the limitation of the present data to $\eta<5.1$. In particular, it yields very large values for the width $\Gamma_{gg}$ (FWHM) of the central source, and correspondingly large pseudorapidity density values at $\eta=y_{\text{beam}}$ which violate the limiting fragmentation scaling hypothesis \cite{alv11}. In order to obtain realistic RDM parameter values, one definitly needs data in the fragmentation region. These will, however, not be available in the next years.

As a remedy, we use the effect of limiting fragmentation scaling \cite{ben69} as a constraint. It has been observed to hold with considerable accuracy in $pp$ and $AA$ collisions \cite{bb03,alv11}: At sufficiently high energy, particle production in the fragmentation region becomes essentially independent of the collision energy. The limiting fragmentation concept refers to particle production as function of rapidity y, but it also holds in pseudorapidity space since for $|\eta|\gg1$ we have $y\simeq\ln{(p_T/m_T)}+\eta\simeq\eta$. (In the large-$\eta$ region, the Jacobian is very close to unity.) 

To supplement the missing LHC data in the pseudorapidty region close to $y_{\text{beam}}$, we consider analogous centrality classes in the 0.2 TeV AuAu results at RHIC, select a small segment of RHIC data (up to five of the outmost datapoints of the PHOBOS datasets), and shift them by
$\Delta y= y_{\text{beam}}^{\text{LHC}}- y_{\text{beam}}^{\text{RHIC}}=7.99-5.36=2.63$. Since the centrality classes of the measurements at RHIC by PHOBOS \cite{bb01} and at LHC by ALICE \cite{to11} do not match exactly for more peripheral collisions, a slight correction of the shift is performed, except for the two most central classes.

The RHIC data points in the fragmentation region are then taken as an additional constraint of our 
$\chi^2-$minimization of the RDM parameters with respect to the preliminary ALICE data at 2.76 TeV.
It turns out that in contrast to an unconstrained $\chi^2-$minimization, this yields physically reasonable distribution functions and RDM parameters, which can be used as basis for an extrapolation to the highest LHC energy of 5.52 TeV. The midrapidity value is mainly determined by the central source at LHC energies, and is therefore not much affected by the use of limiting fragmentation at LHC energies. Small scaling violations which may occur in the forward region would essentially be disconnected from the midrapidity source, although they can slightly modify its weight and width. 

The result of the constrained optimization of the RDM-parameters with respect to the preliminary PbPb data at 2.76 TeV is shown by the solid curve 
in Fig.~\ref{fig1} for central collisions, with parameters given in Table~\ref{tab1}. The integration is 
stopped at the optimum values of 
$\tau_{int}/\tau_{y}$, $\Gamma_{1,2,gg}$, and $n_{gg}$;
the explicit value of $\tau_{int}$ is not needed. The fraction of particles produced in the central source is
$n_{gg}$.  In view of the preliminary character of the ALICE data, we do not list numerical $\chi^2-$results here. 

The dependencies of the parameters on the center-of-mass energy for central collisions is shown in the lhs column in Fig.~\ref{fig2}, and their centrality dependence for 2.76 TeV PbPb in the rhs column of the same figure. Solid curves are fit functions.

The value of the time parameter $\tau_{int}/\tau_{y}$ shown in the upper frame of Fig.~\ref{fig2} is decisive for the position of the fragmentation peak in $\eta-$space. It was found to decrease for increasing $\log\sqrt{s_{NN}}$ from AGS to the highest RHIC energies with a functional dependence discussed in \cite{gw11}, and hence, the extrapolation to LHC energies was based on a continued fall, resulting in $\tau_{int}/\tau_{y}\simeq0.67$ at 2.76 TeV.

The comparison with the preliminary data, however, reveals that it actually levels off at LHC energies to a value of $\tau_{int}/\tau_{y}\simeq0.87$ at 2.76 TeV. This indicates that in the large energy gap between the highest RHIC and the current LHC energy, the rapidity relaxation time $\tau_y$ decreases faster than the interaction time $\tau_{int}$. 

The total number of produced charged hadrons $N_{tot}$, and the fraction $n_{gg}$ produced in the central source are also displayed in Fig.~\ref{fig2}. The latter reaches about 0.56 at 2.76 TeV,
whereas at RHIC energies, it remains below 0.3, such that the multiplicity density at midrapidity has a substantial contribution from the overlapping fragmentation sources \cite{gw11}. 
To decide whether a saturation of $n_{gg}$ is attained at LHC energies, one needs the results from charged-hadron production at the LHC design energy of 5.52 TeV, to be expected in 2014/15. 

The partial widths as functions of energy are displayed in the 
lower two frames of Fig.~\ref{fig2} for both 
fragmentation and midrapidity
sources.
These widths are effective values, because they include the effect of collective expansion in addition to the statistical widths
that can be calculated from a dissipation-fluctuation theorem \cite{wols99}. 

Should future LHC experiments at large pseudorapidity reveal small limiting fragmentation scaling violations, this would
lead to a modification of the widths and weights of the fragmentation sources, and indirectly also to small modifications of the central source.

The values at RHIC energies are resulting from corresponding $\chi^{2}$-minimizations with respect to PHOBOS data \cite{alv11}. Note that RDM parameters at 0.13 and 0.2 TeV (Table~\ref{tab1}) are slightly different from the values in Table 1 of \cite{gw11} which were based on previous data \cite{bb01,bb03}. The data extend in pseudorapidity up to and even sligthly beyond $y_{\text{beam}}$, with a sizeable pseudorapidity density measured at $\eta=y_{\text{beam}}$.

In the rhs column of Fig.~\ref{fig2}, the centrality dependence of the RDM parameters as obtained from the $\chi^2-$minimization including limiting fragmentation is shown for 2.76 TeV PbPb, and seven centrality classes as indicated by the diamonds. The dependences are shown as functions of the average numbers of participants for each centrality class, from $0-5\%$ central to $50-60\%$ peripheral collisions (see also Fig.~\ref{fig3}). For all five RDM parameters, we find a nearly linear dependence on the average number of participants in each class. 

There is only a slight decrease of the time parameter
$\tau_{int}/\tau_{y}$ and the particle fraction in the midrapidity source $n_{gg}$ from central to peripheral collisions. The strongest centrality dependence is found for the total number of produced charged hadrons, which falls from about 17,300 in a $0-5\%$ central collision to below 1,800 in a $50-60\%$ peripheral collision. The relative strength, and the width of the central source decrease only slightly towards peripheral collisions. The width of the fragmentation sources is almost independent of centrality, and slightly smaller than the width of the central source.

The centrality dependence of the pseudorapidity distributions of produced charged hadrons in 2.76 TeV PbPb is displayed in
Fig.~\ref{fig3}, with the dependence of the RDM parameters on the number of participants as displayed in the rhs column of Fig.~\ref{fig2} for seven centralities as indicated in the caption. 

At LHC energies, and for all centralities, the overall scenario changes in favor of particle production from the midrapidity source, as can be seen for central collisions in Fig.~\ref{fig4}, with the $n_{gg}$ values for the fraction of particles in the midrapidity source given in Table~\ref{tab1}. At 2.76 TeV, the bulk of the midrapidity density is generated in the central source,
there is a relatively small overlap of the fragmentation sources at midrapidity. 

The sizeable dip at midrapidity that is seen in the preliminary ALICE data at 2.76 TeV \cite{to11} is likely due to the Jacobian plus the interplay of the three sources, with the fragmentation sources moving much further apart as compared to the highest RHIC energy of 0.2 TeV. The Jacobian transformation has an effect only on the midrapidity source, which is flattened and has a slight dip. In the three-sources approach, the midrapidity minimum seen in the data is then easily achieved.

There exist detailed microscopic calculations of fragmentation sources from 
$gq \rightarrow q$ and $qg \rightarrow q$ diagrams by Szczurek et al. \cite{sz04,csz05} for pion production in proton-proton, and nucleus-nucleus collisions at SPS and RHIC energies. These processes are also responsible for the observed differences \cite{bea01} in the production of positively and negatively charged hadrons, in particular, pions. Extending these calculations to LHC energies would lead to a microscopic foundation of our three-sources approach at energies beyond RHIC.

The existence and relevance of the fragmentation sources is corroborated by results of a partonic model for net-baryon distributions that we had presented and discussed in \cite{mtw09,mtwc09,mtw10}. For baryons minus antibaryons (and also for net protons), the midrapidity gluon-gluon source cancels out, and only the fragmentation sources remain. At high SPS and RHIC energies (and accordingly, in the predictions for LHC), these give rise to two pronounced fragmentation peaks which are clearly seen in the data. Note, however, that the fragmentation peaks in net baryons occur at larger rapidity values as compared to charged hadrons. As an example, the net-baryon peak in 2.76 TeV PbPb ist at $y_{\text{peak}}=5.7$ \cite{mtw11}, whereas the charged-hadron fragmentation peak is at $\eta=3.3$.

With extrapolations of the time parameter in the relativistic diffusion model, the numbers of charged particles in the sources, and the partial widths $\Gamma_{1,2}, \Gamma_{gg}$ from Fig.~\ref{fig2} and Table~\ref{tab1}, a prediction for central PbPb at 5.52 TeV is shown in Fig.~\ref{fig5}. Here the midrapidity value has been extrapolated with $\log{\sqrt{s_{NN}}}$,
and the value of $\Gamma_{gg}$ is determined accordingly. The central collision results for AuAu (0.13 and 0.2 TeV) and PbPb at 2.76 TeV 
are displayed for comparison. 


\section{Conclusion}
We have analyzed recent preliminary ALICE results for PbPb collisions at LHC energy of $\sqrt{s_{NN}}$ = 2.76 TeV. Charged-hadron pseudorapidity distributions have been calculated analytically in the
non-equilibrium statistical relativistic diffusion model RDM. For seven different centralities, the underlying RDM parameters have been determined in a $\chi^2$-optimization of the analytical model solutions with respect to the preliminary data, and using limiting fragmentation as an additional constraint. 

A comparison with a previous prediction \cite{gw11} that was based on an extrapolation of the parameters with $\log\sqrt{s_{NN}}$ reveals that the rapidity relaxation time $\tau_y$ decreases substantially in the energy region between RHIC and LHC energies, leading to a larger time parameter $\tau_{int}/\tau_y$ and hence, to a fragmentation-peak position that is closer to midrapidity than expected from the earlier extrapolation of the time parameter.


Based on the RDM fit to the data in the three-sources model, the midrapidity source that is associated with gluon-gluon collisions accounts for about 56\% of the total charged-particle multiplicity measured by ALICE in central PbPb collisions at 2.76 TeV. 

The fragmentation sources that correspond to particles that are mainly generated from valence quark -- gluon interactions are centered at pseudorapidity values $<\eta_{1,2}>\simeq <y_{1,2}> \simeq \mp 3.3$. The total particle content in these sources amounts to about 44\% of the total charged hadron production, but contributes only marginally to the midrapidity yield. 
It is, however, decisive for the more pronounced midrapidity dip at LHC energies where the fragmentation sources move much further apart than at RHIC energies. 

With the results for PbPb at 2.76 TeV LHC energy  and previous RDM results for AuAu collisions at RHIC energies, we have extrapolated the three-sources model parameters to the LHC design energy of 5.52 TeV for PbPb, and calculated the corresponding charged-hadron pseudorapidity distribution.
Small corrections of the extrapolated values for the diffusion-model parameters may be required once the final measured distributions become available at both LHC energies.\\\\

This work has been supported by the ExtreMe Matter Institute EMMI.
\bibliography{gw_prc_nt}

\end{document}